\documentclass[aps,pra,twocolumn,showpacs,superscriptaddress]{revtex4-1}

\usepackage{graphicx}
\usepackage[usenames]{color}
\usepackage{amssymb,amsmath}
\usepackage{leftidx}

\newcommand{\eq}[1]{Eq.~(\ref{#1})}
\newcommand{\fig}[1]{Fig.~\ref{#1}}
\newcommand{\be}[1]{\begin{equation}\label{#1}}
\newcommand{\ee}{\end{equation}}

\begin{document}

\title{Traces on  ion yields and electron spectra of Ar inner-shell hollow states  with Free-Electron Lasers  }

\author{A. O. G. Wallis}
\affiliation{Department of Physics and Astronomy, University College London, Gower Street, London WC1E 6BT, United Kingdom}
\author{H. Banks}
\affiliation{Department of Physics and Astronomy, University College London, Gower Street, London WC1E 6BT, United Kingdom}
\author{A. Emmanouilidou}
\affiliation{Department of Physics and Astronomy, University College London, Gower Street, London WC1E 6BT, United Kingdom}

\begin{abstract}
We explore the formation by Free-Electron-Laser radiation of Ar hollow states with two or three inner-shell holes. We find that even charged Ar ion states 
 can be more populated than odd charged Ar ion states. This depends on the pulse intensity and the number of energetically accessible inner-shell holes. Fully accounting for fine structure, we demonstrate that one electron spectra bare the imprints of Ar hollow states with two inner-shell holes. Moreover, we show how the Auger spectra of these hollow states can be extracted from two-electron coincidence spectra.

\end{abstract}


\maketitle

The advent of  extreme ultraviolet and X-ray Free-Electron Lasers (FELs) allows the exploration of novel states of matter. One fascinating aspect of FELs is that the laser boils away electrons from the inside out giving rise to hollow atoms and molecules.  To monitor the femtosecond time-scale dynamics of these hollow states one needs to identify 
the ionization pathways that lead to their formation. Understanding the processes leading to the formation of hollow states will allow 
these states to be employed as the basis for a new type 
of spectroscopy for chemical analysis \cite{spectroscopy1,spectroscopy2,spectroscopy3,spectroscopy4}. It will also assist in achieving atomic resolution in diffraction patterns from biological molecules interacting with FEL-radiation \cite{application1,application2}.

We consider Ar interacting with FEL radiation. For each additional inner-shell hole that becomes energetically accessible, a link of a  $\mathrm{P_{C}}$ and an $\mathrm{A_{V}}$ transition is added to the ionization pathways. Even charged Ar ion states are primarily populated by chains of these links.
 $\mathrm{P}$  stands for a single-photon ionization of an electron and $\mathrm{A}$ for an Auger decay with an electron from a higher orbital, denoted as the subscript in A, dropping to fill in a hole. $\mathrm{C}$ and $\mathrm{V}$ stand for a core and a valence electron, respectively. We show that when hollow states with two inner-shell holes are formed, the ion yield of $\mathrm{Ar^{2n+}}$  is larger than the ion yield of  $\mathrm{Ar^{(2n-1)+}}$, with $\mathrm{n=1, ..., h}$ and $\mathrm{h}$ the number of holes. This is true for all intensities. However, when three inner-shell holes are energetically accessible, additional transitions become available. These are Coster-Kronig Auger ($\mathrm{A_C}$) transitions \cite{Coster} where the hole and the electron dropping in to fill the hole occupy sub-shells with the same $\mathrm{n}$ and  different $\mathrm{l}$  numbers. As a result, we find that it is only for higher intensities that the yield of the even charged Ar ion states is larger than the yield of the odd charged Ar ion states.

Focusing on two inner-shell holes, we demonstrate how to identify the formation of the $\mathrm{Ar^{2+}(2p^{-2})}$ hollow state \cite{covmapping, moleculehollowBerrah,atomshollow}. We show that the yield of $\mathrm{Ar^{4+}}$, which bares the imprint of  $\mathrm{Ar^{2+}(2p^{-2})}$, is not sufficient for identifying this hollow state.  The reason is that $\mathrm{Ar^{4+}}$ is populated by competing ionization pathways, however, not all of these pathways contribute to the formation of the hollow state. Unlike in the ion yields, we find that in the one electron spectra these competing pathways leave different traces and we can thus discern the formation  of  $\mathrm{Ar^{2+}(2p^{-2})}$. We also show how to extract the Auger spectrum of the hollow state from two-electron coincidence spectra.

We first describe the rate equations  we use to obtain our results \cite{Alisdair1,rates}. We account for the general case  when multiple states lead to state $\mathrm{j}$, for example, $\mathrm{i \rightarrow j \rightarrow k}$ and $\mathrm{i' \rightarrow j \rightarrow k}$.
To compute the contribution of the state $\mathrm{i}$ to the yield $\mathrm{\mathcal{I}^{(q-1)}_{j(i)}}$ of the ion state $\mathrm{j}$ with charge $\mathrm{q-1}$ we solve the rate equations:
\begin{align}\label{eq:rate2}
\mathrm{\frac{d}{dt} \mathcal{I}^{(q-1)}_{j(i)}(t)=}
& \mathrm{(\sigma_{i\rightarrow j}J(t)+\Gamma_{i\rightarrow j})\mathcal{I}^{(q-2)}_{i}(t)} \\ \nonumber 
&\mathrm{- \sum_{k'}(\sigma_{j\rightarrow k'}J(t)+\Gamma_{j\rightarrow k'})\mathcal{I}^{(q-1)}_{j(i)}(t)} \\ \nonumber
\mathrm{\frac{d}{dt} \mathcal{P}^{(q)}_{j(i) \rightarrow k} =} &\mathrm{\sigma_{j\rightarrow k}J(t)\mathcal{I}^{(q-1)}_{j(i)}(t)} \\ \nonumber
\mathrm{\frac{d}{dt} \mathcal{A}^{(q)}_{j(i) \rightarrow k} }= & \mathrm{\Gamma_{j\rightarrow k}\mathcal{I}^{(q-1)}_{j(i)}(t)}, \end{align}
where $\mathrm{\sigma_{i\rightarrow j}}$ and $\mathrm{\Gamma_{i\rightarrow j}}$  are the single-photon absorption cross section and the Auger decay rate from the initial state $\mathrm{i}$ to the final state $\mathrm{j}$, respectively. $\mathrm{J(t)}$ is the photon flux, which is modeled with a Gaussian function. Atomic units are used in this work. For details on how we compute $\mathrm{\Gamma_{i\rightarrow j}}$, see \cite{Alisdair1}.
The first term in \eq{eq:rate2} accounts for the formation of the state $\mathrm{j}$ with charge $\mathrm{q-1}$ through the single-photon ionization and the Auger decay of the state $\mathrm{i}$ with charge $\mathrm{q-2}$. The second term in \eq{eq:rate2}  
 accounts for the depletion of state $\mathrm{j}$ by single-photon ionization and  Auger decay to the state $\mathrm{k'}$ with charge $\mathrm{q}$.
In addition, we compute the photo-ionization $\mathrm{\mathcal{P}^{(q)}_{j(i) \rightarrow k}}$ and the Auger $\mathrm{\mathcal{A}^{(q)}_{j(i) \rightarrow k}}$ yields, with  $\mathrm{q}$ the charge of the final state $\mathrm{k}$.  These yields provide the probability for observing two electrons with energies corresponding to the transitions $\mathrm{i\rightarrow j}$ and  $\mathrm{j\rightarrow k}$. Using  these yields, we obtain the coincidence two-electron spectra. The one electron spectra, that is, the transition yields  from an initial state $\mathrm{j}$ with charge $\mathrm{q-1}$ to a final state $\mathrm{k}$ with charge $\mathrm{q}$ and the  ion yields of the state  $\mathrm{j}$ and of all the states  with  charge $\mathrm{q-1}$ are given by
\begin{align}\label{eqn:rate3}
&\mathrm{\mathcal{P}^{(q)}_{j\rightarrow k} }= \mathrm{\sum_{i} \mathcal{P}^{(q)}_{j(i) \rightarrow k}\quad\mathcal{A}^{(q)}_{j\rightarrow k}} =\mathrm{ \sum_{i} \mathcal{A}^{(q)}_{j(i) \rightarrow k}}\\
&\mathrm{\mathcal{I}^{(q-1)}_{j} }= \mathrm{\sum_{i} \mathcal{I}^{(q-1)}_{j(i)}}
\quad\mathrm{\mathcal{I}^{(q-1)} }= \mathrm{\sum_{j} \mathcal{I}^{(q-1)}_{j}}.
\end{align}

\begin{figure}
\begin{center}
\includegraphics[width=0.8\linewidth]{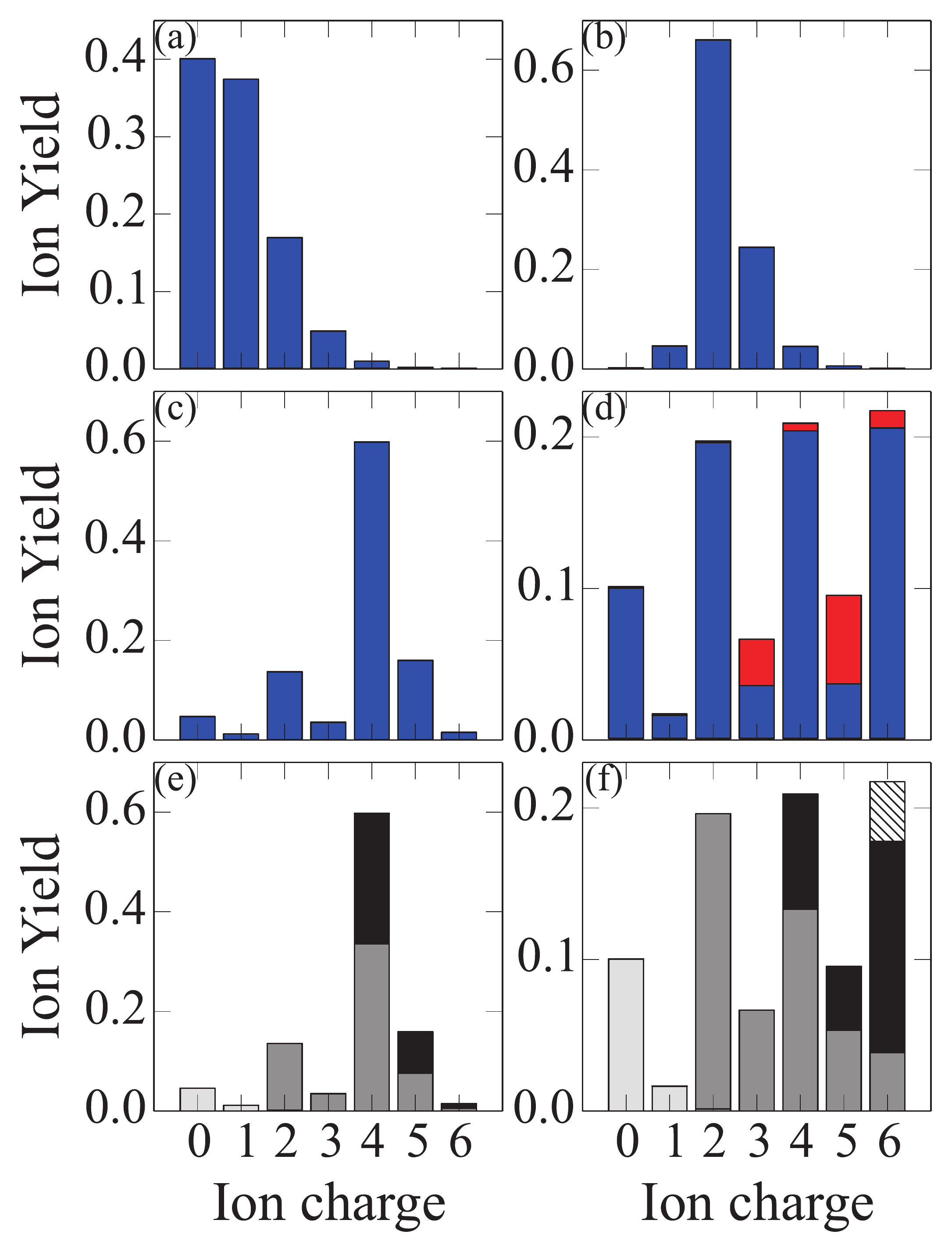}
\caption{
Ion yields of $\mathrm{Ar^{n+}}$ for a pulse of 5$\times10^{15}$ W cm$^{-2}$ intensity, 10 fs duration and different photon energies. For each photon energy, the number of accessible inner-shell holes is different:
(a) 200 eV, no inner-shell holes;
(b) 260 eV, a single $\mathrm{2p}$ inner-shell hole;
(c) 315 eV, two $\mathrm{2p}$ inner-shell holes. 
(d) 360 eV,  three $\mathrm{2p}$ and a combination of two $\mathrm{2p}$ and one $\mathrm{2s}$ inner-shell  holes. Highlighted in red is the contribution of Coster-Kronig Auger transitions. (e) for 315 eV  and (f) for 360 eV show the contribution of pathways 
 that are differentiated by the maximum number of core holes in any state along each pathway: light grey corresponds to zero maximum number of core holes, grey to one, black to two and striped black lines to three.}
\label{fig:coreholes}
\end{center}
\end{figure}

\begin{figure}[h]
\begin{center}
\includegraphics[width=0.8\linewidth]{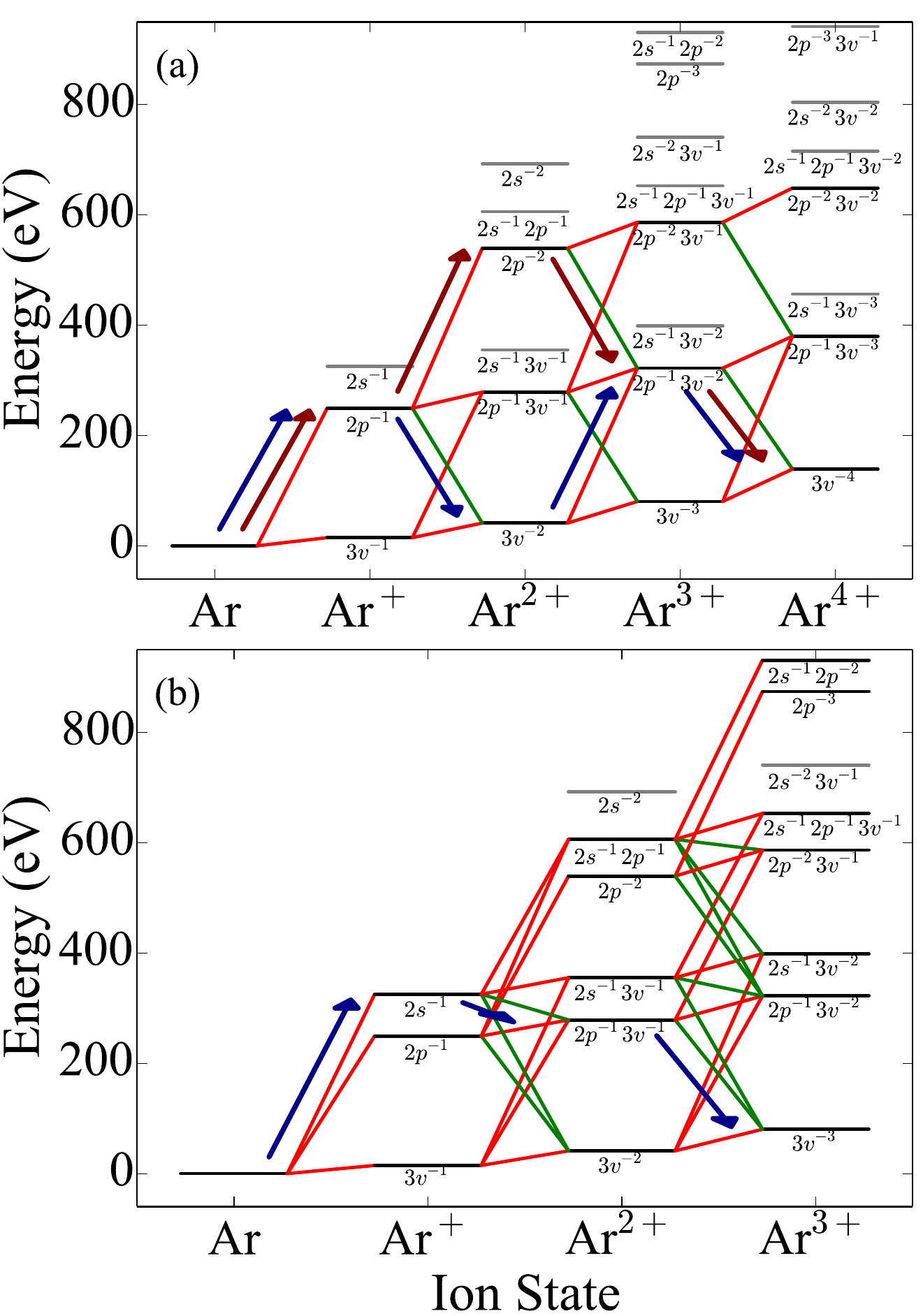}
\caption{  Ionization pathways between different electronic configurations of Ar accessible with $\mathrm{P}$ (red lines) and $\mathrm{A}$ (green lines) events (a) up to Ar$^{4+}$ for  $\mathrm{\hbar \omega=315}$ eV and (b) up to $\mathrm{Ar^{3+}}$ for  $\mathrm{\hbar \omega=360}$ eV.  
The labels $\mathrm{2s^{-a}2p^{-b}}3v^{-c}$ stand for the electronic configuration $\mathrm{(2s^{2-a}2p^{6-b}3s^{2-d}3p^{6-e})}$, with $\mathrm{d+e=c}$ the number of valence holes. In (a) $\mathrm{P_{C}A_{V}P_{C}A_{V}}$ (blue arrows) and $\mathrm{P_{C}P_{C}A_{V}A_{V}}$ (brown arrows) are the pathways which contribute the most to  the ion yield of $\mathrm{Ar^{4+}}$. In (b) the blue arrows indicate the pathway $\mathrm{P_{C}A_{C}A_{V}}$ that involves  a Coster-Kronig $\mathrm{A_{C}}$  transition  and populates  $\mathrm{Ar^{3+}}$.}
\label{fig:energylevels}
\end{center}
\end{figure}

In \fig{fig:coreholes}  we compute the yields of the $\mathrm{Ar^{n+}}$ ion states for four photon energies and for a high pulse intensity of 5$\times10^{15}$ W cm$^{-2}$. We do so accounting only for the electronic configuration of the ion states in the rate equations and without including fine structure \cite{Alisdair1}. We first consider  a photon energy sufficiently low, 200 eV, that single-photon ionization events do not lead to the formation of inner-shell holes.  In \fig{fig:coreholes}(a), we show  that  the  $\mathrm{Ar^{n+}}$ ion states are populated in descending order. For 260 eV, a single $\mathrm{2p}$ inner-shell hole is accessible by a $\mathrm{P_C}$ process
from neutral Ar. 
 A $\mathrm{P_C}$ is a much more likely transition than a  $\mathrm{P_V}$ one. As a result, for all intensities, the population going through $\mathrm{Ar^{+}(2p^{-1})}$ is much larger than the population ending up in or going through $\mathrm{Ar^{+}(3}v^{-1})$. In addition, since the $\mathrm{P_C}$ photo-ionization is followed by an $\mathrm{A_{V}}$ decay---$\mathrm{P_CA_V}$ pathway---the ion yield of $\mathrm{Ar^{2+}}$ is higher than the ion yield of $\mathrm{Ar^{+}}$, see \fig{fig:coreholes}(b).
For 315 eV, two $\mathrm{2p}$ inner-shell holes are accessible by two $\mathrm{P_C}$ events, see \fig{fig:energylevels}(a). As for 260 eV, $\mathrm{Ar^{2+}}$ has a larger population than $\mathrm{Ar^{+}}$. In addition,  $\mathrm{P_CA_{V}P_CA_{V}}$ and $\mathrm{P_CP_CA_{V}A_{V}}$ are now energetically allowed pathways that populate $\mathrm{Ar^{4+}}$, see \fig{fig:energylevels}(a). Since, $\mathrm{Ar^{3+}}$ is populated by pathways involving at least one $\mathrm{P_V}$ process, $\mathrm{Ar^{4+}}$ has a larger population than $\mathrm{Ar^{3+}}$, see  \fig{fig:coreholes}(c).  For 360 eV, three $\mathrm{2p}$ inner-shell holes or a combination of one $\mathrm{2s}$ and two $\mathrm{2p}$ inner-shell holes are accessible through three $\mathrm{P_C}$ events, see \fig{fig:energylevels}(b). Pathways involving three $\mathrm{P_C}$ and three $\mathrm{A_V}$ transitions, such as  $\mathrm{P_CA_{V}P_CA_{V}P_CA_{V}}$, are now energetically allowed and populate $\mathrm{Ar^{6+}}$. For 200 eV, 260 eV and 315 eV the odd charged states are populated only by pathways that include at least one $\mathrm{P_V}$ process. In contrast, for 360 eV, pathways that include Coster-Kronig Auger transitions  between the $\mathrm{2s}$ and $\mathrm{2p}$ sub-shells are energetically allowed. These pathways do not necessarily involve a $\mathrm{P_V}$ event. For instance, in \fig{fig:energylevels}(b), we show the $\mathrm{P_CA_{C}A_V}$ pathway that includes a Coster-Kronig transition ($\mathrm{A_C}$) and populates 
 $\mathrm{Ar^{3+}}$. When no Coster-Kronig transitions are present,  the most probable pathways populating the $\mathrm{Ar^{(2n-1)+}}$ states and those populating the $\mathrm{Ar^{(2n)+}}$ states have the same number of P events, with $\mathrm{n=1,...,h}$. Thus, for 260 eV and  for 315 eV, the yield of the Ar$^{(2n-1)+}$ states is less than the yield of the Ar$^{(2n)+}$ states for all intensities. However, when a Coster-Kronig transition is present some of the most probable pathways populating the $\mathrm{Ar^{(2n-1)+}}$ states  have one P transition less than the most probable pathways populating the $\mathrm{Ar^{(2n)+}}$ states. As a result, the yield of the $\mathrm{Ar^{(2n-1)+}}$ states  is larger/smaller than the yield of the Ar$^{(2n)+}$ states for low/high intensities.  For a high pulse intensity of 5$\times10^{15}$ W cm$^{-2}$, in \fig{fig:coreholes}(d), we show that the ion yields of $\mathrm{Ar^{4+}}$ and $\mathrm{Ar^{6+}}$  are larger than the ion yields of $\mathrm{Ar^{3+}}$ and $\mathrm{Ar^{5+}}$, respectively.

For the results in \fig{fig:coreholes}, double ionization  (DI) and double Auger (DA) decays are not accounted for. These are both processes where two electrons are ejected in one step. Some of the pathways DI and DA give rise to have one P process less compared to pathways where only one electron is ejected at each ionization step. As a result,  the contribution of these two processes is less for high intensities. Moreover, these two processes are significantly less likely than the ionization processes we currently account for in our calculations. For instance,  the probability for a DA decay from a 2p hole in Ar is roughly 10\% of the probability for a single Auger decay \cite{doubleAuger}.


\begin{figure}
\begin{center}
\includegraphics[width=0.5\textwidth]{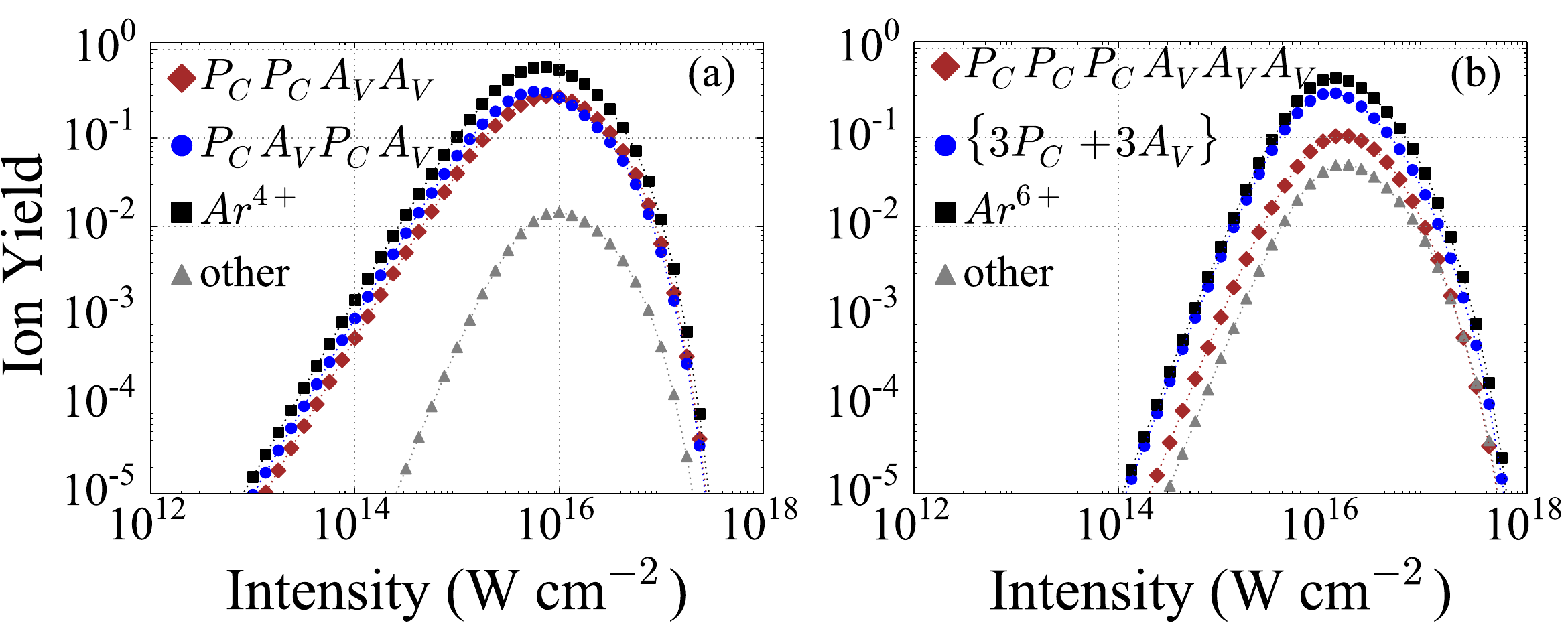}
\end{center}
\caption{As a function of pulse intensity: (a) for 315 eV  and 10 fs, the ion yields of $\mathrm{Ar^{4+}}$  (black squares), of $\mathrm{P_CA_{V}P_CA_{V}}$  (blue circles), of $\mathrm{P_CP_{C}A_VA_{V}}$  (brown diamonds) and of all the other pathways contributing to $\mathrm{Ar^{4+}}$ (grey triangles); (b) for  360 eV and 10 fs, the ion yield of $\mathrm{Ar^{6+}}$ (black squares), of $\mathrm{P_CP_{C}P_CA_{V}A_VA_V}$  (brown diamonds),  of  \{$\mathrm{3P_{C}+3A_V}$\} (blue circles) and of all the other pathways contributing to $\mathrm{Ar^{6+}}$ (grey triangles). 
\label{fig:hollowstates}}
\end{figure}

Focusing on hollow states with two inner-shell holes, we look for observables with clear  imprints of $\mathrm{Ar^{2+}(2p^{-2})}$. We first consider the ion yields. In \fig{fig:hollowstates}(a), for 315 eV, we plot as a function of intensity the ion yield of $\mathrm{Ar^{4+}}$. We also plot  the contributions to this latter yield of the $\mathrm{P_CP_CA_{V}A_{V}}$ and of the  $\mathrm{P_CA_VP_{C}A_{V}}$ pathways and the contribution of all the remaining pathways. $\mathrm{Ar^{2+}(2p^{-2})}$ is formed from Ar by two sequential $\mathrm{P_C}$ events while it is  is depleted through two sequential $\mathrm{A_V}$ events. Thus, the $\mathrm{P_CP_CA_{V}A_{V}}$ pathway bares the imprint of the formation of $\mathrm{Ar^{2+}(2p^{-2})}$, see \fig{fig:energylevels}(a).
 In \fig{fig:hollowstates}(a), we also show that the yield of all other pathways that involve a $\mathrm{P_V}$ transition  is much smaller than the yields of $\mathrm{P_CP_CA_{V}A_{V}}$ and $\mathrm{P_CA_VP_{C}A_{V}}$. We choose a small pulse duration, 10 fs, since
 it favors the contribution of the $\mathrm{P_CP_CA_{V}A_{V}}$ pathway. The reason is that for small pulse durations high intensities are reached faster. This  favors a $\mathrm{P_CP_C}$ sequence rather than a $\mathrm{P_CA_V}$ sequence. However, even for this small pulse duration, the yield of $\mathrm{P_CP_CA_{V}A_{V}}$ is  similar to the yield  of $\mathrm{P_CA_VP_{C}A_{V}}$. \fig{fig:coreholes}(e) also shows that ion yields alone do not trace the formation of  $\mathrm{Ar^{2+}(2p^{-2})}$. Indeed, pathways that go through the two inner-shell hollow state contribute to the ion yield of $\mathrm{Ar^{4+}}$ as much, if not less, as pathways that go through hollow states with up to one inner-shell hole.
For hollow states with three inner-shell holes, it is even more difficult to discern the pathway that bares the imprint of the hollow state.  We show this to be the case  for 360 eV in \fig{fig:hollowstates}(b) where we plot as a function of intensity the yield of $\mathrm{Ar^{6+}}$. We also plot the contribution to this latter yield of the $\mathrm{P_CP_CP_CA_{V}A_{V}A_V}$ pathway, the contribution of the sum of the other \{$\mathrm{3P_{C}+3A_V}$\} pathways that involve three $\mathrm{P_C}$ and three $\mathrm{A_V}$ events and the contribution of all the remaining pathways. The $\mathrm{P_CP_CP_CA_{V}A_{V}A_V}$ pathway bares the imprint of the $\mathrm{Ar^{3+}(2p^{-2}2s^{-1})}$ and of the $\mathrm{Ar^{3+}(2p^{-3})}$ states.  Its yield is smaller than the yield of the other \{$\mathrm{3P_{C}+3A_V}$\} pathways. Also, it is only slightly larger than the sum of the yields of the pathways that involve at least one $\mathrm{P_V}$ process. \fig{fig:coreholes}(f) also shows that, using ion yields alone, we can not discern the formation of the states $\mathrm{Ar^{3+}(2p^{-2}2s^{-1})}$ and  $\mathrm{Ar^{3+}(2p^{-3})}$. Indeed, the largest contribution to the ion yield of  $\mathrm{Ar^{6+}}$ comes from pathways that go through states with up to  two inner-shell holes.  

\begin{figure}[h]
\begin{center}
\includegraphics[width=0.9\linewidth]{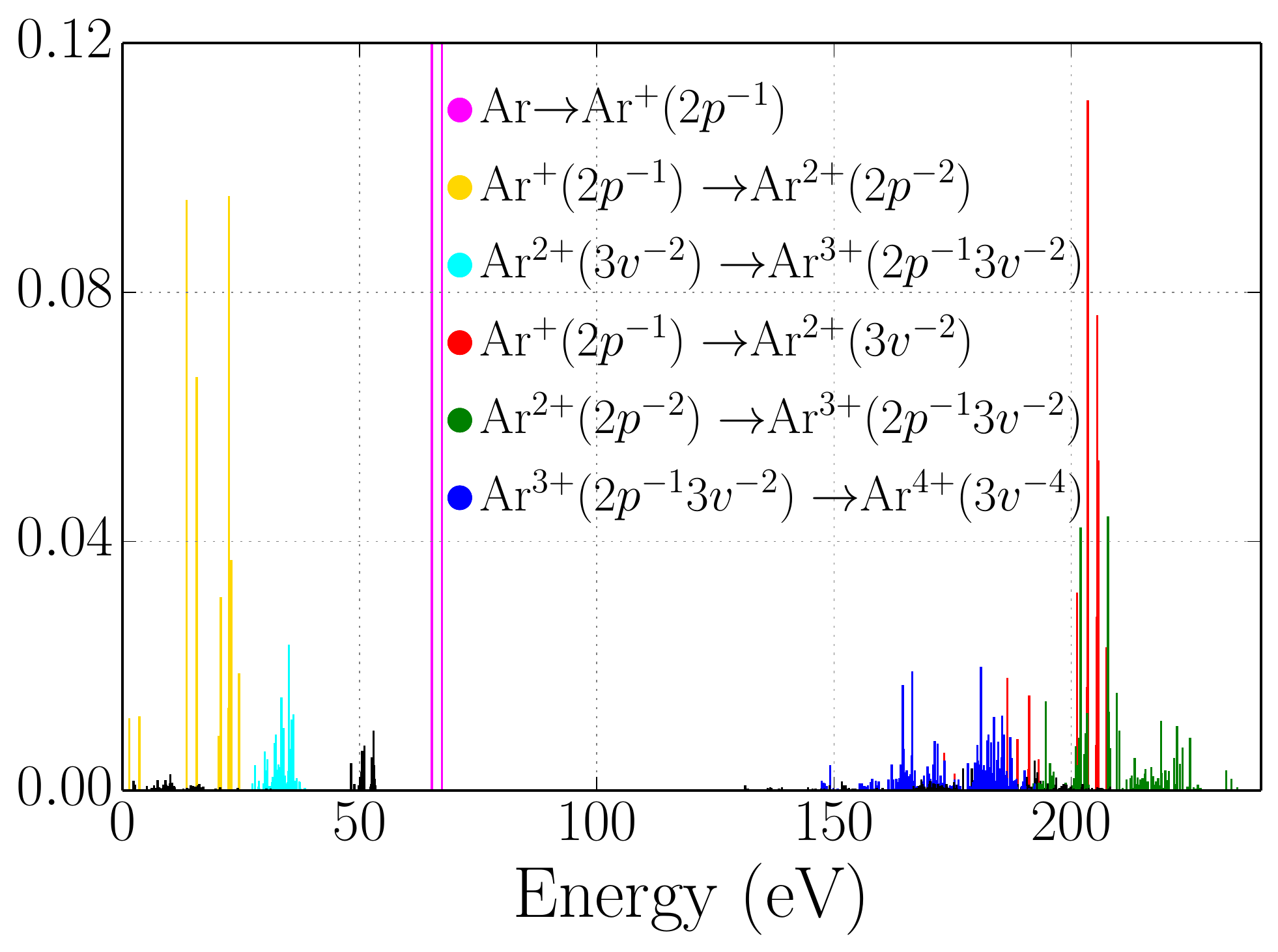}
\caption{
One electron spectra, for  a  pulse of $5\times10^{15}$ W cm$^{-2}$ intensity, 10 fs duration and 315 eV photon energy.}
\label{fig:augerspectra1}
\end{center}
\end{figure}

 We now explore whether we can identify the formation of $\mathrm{Ar^{2+}(2p^{-2})}$ from one electron spectra.  In \fig{fig:augerspectra1}, for 315 eV, we compute the one electron photo-ionization and Auger spectra  for pulse parameters that optimize the contribution of the $\mathrm{P_CP_CA_{V}A_{V}}$ pathway.  Unlike the ion yields, to accurately calculate the electron spectra we now fully account for the fine structure of the ion states in the rate equations.   To obtain these fine structure ion states we  perform calculations using the \textsc{grasp2k} \cite{grasp2k:2013} and \textsc{ratip} \cite{RATIP:2012} packages, within the relativistic Multi-Configuration Dirac-Hartree-Fock (MCDHF) formalism, see \cite{Alisdair1} for details.   Out of all $\mathrm{P}$ and $\mathrm{A}$ transitions  shown in \fig{fig:energylevels}(a), the transitions that bare the imprint of $\mathrm{Ar^{2+}(2p^{-2})}$ are the $\mathrm{P_C}$ transition   $\mathrm{Ar^+(2p^{-1})\rightarrow Ar^{2+}(2p^{-2})}$ and the $\mathrm{A_V}$
 transition 
 $\mathrm{Ar^{2+}(2p^{-2})\rightarrow Ar^{3+}(2p^{-1}}3v^{-2})$. Can we separate these transitions from all others in one electron spectra?
 
For 315 eV, the single-photon ionized electrons from an inner-shell (2s or 2p) escape with energies between 0 eV and 70 eV while those ionized   from a valence shell (3s or 3p) escape with energy between 214 eV and 300 eV. The $\mathrm{\mathcal{P}^{(q)}_{j \rightarrow k}}$ yields for valence shell electrons  are  very small and not visible 
 in \fig{fig:augerspectra1}. Moreover, the Auger electrons escape with energies between 150 eV and 240 eV. Thus, Auger electrons  are well separated from single-photon ionized electrons.
 The most probable Auger and single-photon ionization transitions are depicted in \fig{fig:augerspectra1}. During the transitions: i)   
$\mathrm{Ar^+(2p^{-1})\rightarrow Ar^{2+}}(3v^{-2})$ the first Auger electron is ejected along $\mathrm{P_CA_VP_{C}A_{V}}$ with energy from 173 eV to 208 eV \cite{grasp2k:2013}, ii)  $\mathrm{Ar^{2+}(2p^{-2})\rightarrow Ar^{3+}(2p^{-1}}3v^{-2})$  the first Auger electron is ejected along $\mathrm{P_CP_CA_{V}A_{V}}$  with energy from 181 eV to 241 eV and iii) $\mathrm{Ar^{3+}(2p^{-1}}3v^{-2})\mathrm{\rightarrow Ar^{4+}}(3v^{-4})$ the second Auger electron is ejected  along $\mathrm{P_CP_CA_{V}A_{V}}$ and $\mathrm{P_CA_VP_{C}A_{V}}$  with energy from 140 eV to 198 eV. Thus, the Auger transition ii) that bares the imprint of $\mathrm{Ar^{2+}(2p^{-2})}$ can be clearly discerned only for energies above 208 eV. For smaller energies Auger transitions i) and ii) strongly overlap. During the transitions: iv) $\mathrm{Ar\rightarrow Ar^{+}(2p^{-1})}$ the first photo-ionized electron is ejected along  
$\mathrm{P_CP_CA_{V}A_{V}}$ and $\mathrm{P_CA_VP_{C}A_{V}}$ with energy from 65 eV to 67.5 eV, v) $\mathrm{Ar^{+}(2p^{-1})\rightarrow Ar^{2+}(2p^{-2})}$ the second photo-ionized electron is ejected along 
$\mathrm{P_CP_CA_{V}A_{V}}$  with energy from 1 eV to 25 eV and vi) $\mathrm{Ar^{2+}}(3v^{-2})\mathrm{\rightarrow Ar^{3+}(2p^{-1}}3v^{-2})$ the second photo-ionized electron is ejected along 
 $\mathrm{P_CA_VP_{C}A_{V}}$ with energy from 22 eV to 41 eV. From 22 eV to 25 eV there is an overlap between photo-ionization transitions v) and vi). However, transition v) is orders of magnitude larger than vi). Since transition v) bares the imprint of $\mathrm{Ar^{2+}(2p^{-2})}$,   we can clearly identify the formation of the hollow state from the one-electron spectra.

\begin{figure}[h]
\begin{center}
\includegraphics[width=0.9\linewidth]{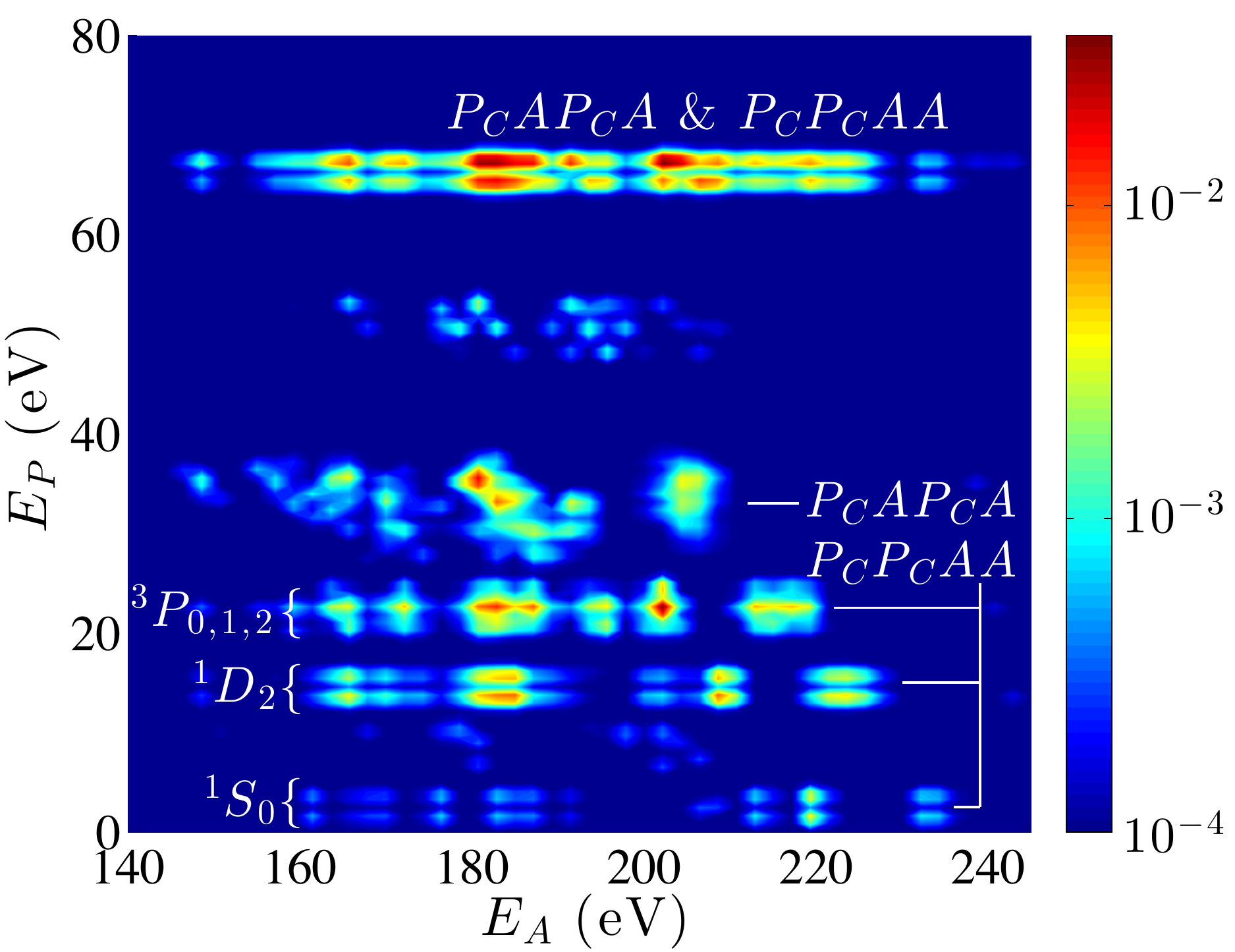}
\end{center}
\caption{Coincidence spectra of an Auger and a photo-ionized electron. The pulse parameters are $5\times10^{15}$ W cm$^{-2}$ intensity, 10 fs duration and 315 eV photon energy.} 
\label{fig:augerspectra2}
\end{figure}

 Finally, we show    how to extract the Auger spectrum of  $\mathrm{Ar^{2+}(2p^{-2})}$  from two-electron coincidence spectra. 
 Coincidence experiments have been performed extensively with synchrotron radiation \cite{Lablanquie:multi:2011,Huttula:2013}. It is expected that coincidence experiments with FEL-radiation will  take place in the near future   \cite{Rudenko,covmapping}. 
 In anticipation of these experiments, in \fig{fig:augerspectra2}, we plot the coincidence spectra of a single-photon ionized electron and an Auger electron. This choice of electrons is based on the fact that single-photon ionized electrons are well separated in energy from Auger electrons, see \fig{fig:augerspectra1}.
  Moreover, we have already shown that for energies of a single-photon ionized electron ($\mathrm{E_{P}}$) up to 25 eV we can clearly discern the second $\mathrm{P_C}$ event---previously denoted as transition v)---in  the $\mathrm{P_CP_CA_{V}A_{V}}$ pathway. Indeed, as shown in \fig{fig:augerspectra2}, there is no trace of the     $\mathrm{P_CA_VP_{C}A_{V}}$ pathway for $\mathrm{E_P <25}$ eV. Focusing on $\mathrm{E_P <25}$ eV, the energies of the Auger electron from 140 eV to 198 eV  correspond to the  transition $\mathrm{Ar^{3+}(2p^{-1}}3v\mathrm{^{-2})\rightarrow Ar^{4+}(}3v^{-4})$, while from 181 eV to 241 eV correspond to the transition $\mathrm{Ar^{2+}(2p^{-2})\rightarrow Ar^{3+}(2p^{-1}}3v^{-2})$. It is this latter transition that corresponds to the Auger spectra of the $\mathrm{Ar^{2+}(2p^{-2})}$ hollow state.
  In more detail, 
  the Auger spectrum of the $\mathrm{^1{S}_{0}}$ fine structure state is the sum of the spectra corresponding to $\mathrm{E_P}$  around 1.4 eV and 3.6 eV. These two energies correspond to the 
  $\mathrm{^2{P}_{3/2}}$ and  $\mathrm{^2{P}_{1/2}}$ fine structure states of $\mathrm{Ar^{+}(2p^{-1})}$. The Auger spectrum of the $\mathrm{^1D_{2}}$ fine structure state is the sum of the spectra corresponding to $\mathrm{E_P}$ around  
  13.5 eV and 15.6 eV. Finally, it is more difficult to discern the Auger spectra of the  $\mathrm{^3P_{0,1,2}}$ fine structure states  in the interval 20.2 eV$\mathrm{< E_P <}$ 24.6 eV. It is also mainly the Auger spectra of these  $\mathrm{^3P_{0,1,2}}$ states that overlaps in the energy interval from 181 eV to 198 eV with the Auger transition $\mathrm{Ar^{3+}(2p^{-1}}3v\mathrm{^{-2})\rightarrow Ar^{4+}(}3v^{-4})$.

  In conclusion, we explored how  Ar states with  multiple inner-shell holes affect the ion yields. We found that the ion yields of even charged ion states are larger than the ion yields of odd charged ion states either for all intensities or only for higher ones. This depends on the type of transitions that are energetically allowed. Our results hold for two and three inner-shell holes in Ar. It would be interesting to further explore how our results are affected by an even larger number of inner-shell holes. Finally, motivating future FEL coincidence experiments, 
  we demonstrated how  two-electron spectra carry information regarding the Auger spectra of hollow states.

A.E. acknowledges support from EPSRC under Grant No. H0031771 and J0171831  and use of the Legion computational resources at UCL.


\end{document}